\documentclass[prl,twocolumn,showpacs]{revtex4}
\usepackage{graphicx}
\begin{document}

\preprint{}

\title{The dynamical Casimir effect and energetic sources for gamma ray bursts}

\author{She-Sheng Xue}

\email{xue@icra.it}

\affiliation{ICRA, INFN and
Physics Department, University of Rome ``La Sapienza", 00185 Rome, Italy}



\begin{abstract}
On the analogy of the Casimir effect, we present an effect of quantum-field fluctuations, 
attributed to gravitational field coupling to the zero-point energy
of virtual particles in the vacuum. In the process of black hole's formation, such an effect could cause tremendous energy release, possibly describing a scenario of energetic sources for 
observed gamma ray bursts.

\end{abstract}

\pacs{04.62.+v, 04.70.Bw, 04.70.Dy}

\maketitle

The mystery of energetic sources generating gamma ray bursts,  
a prompt emission of extremely huge energy, has stimulated many 
studies in connection with electromagnetic 
properties of black holes\cite{dr,prx,putten,koide}. Various astrophysical scenarios are
discussed in literatures\cite{piran}. We present an alternative idea and  
scenario on the basis of gravitational field interacting with the vacuum energy 
(zero-point energy) of virtual 
particles in the vacuum. Studying the vacuum energy in 
the Schwarzschild geometry around a gravitationally 
collapsing matter, we show that the vacuum gains tremendous energy from gravitational field
in the process of black hole's formation, which causes the vacuum becomes unstable, leading to 
energy release.
 
In quantum field theories, the vacuum is composed from virtual particles whose energy
(the vacuum energy or zero-point energy) ${\cal E}$ does not vanish.
The Casimir effect\cite{casimir} shows that the vacuum energy ${\cal E}$ is modified by boundary 
conditions and the vacuum gains the Casimir energy,
\begin{equation}
\delta{\cal E}_c =-{\pi^2\over720 a^3}<0,
\label{casimir}
\end{equation} 
where $a$ is the distance between two plates. Thus the vacuum becomes energetically unstable and 
has to quantum-mechanically decay. This leads to releasing the Casimir energy, and as a result, 
an attractive force between two plates is observed as the Casimir effect.   

In this short letter
, we consider how the vacuum energy ${\cal E}$ is 
modified by the gravitational field around a gravitationally
collapsing matter $M$, described by the Schwarzschild geometry,
\begin{equation}
ds^2\!=\!-\!g(r)dt^2\!+\!g^{\!-\!1}(r)dr^2\!+\!r^2d\Omega,\hskip0.1cm
g(r)\!\equiv\! (1\!-\!{2M\over r})
\label{sg}
\end{equation}
where $\Omega$ is the spherical solid angle and $r,\theta,\phi,t$
are the Schwarzschild coordinates. 

Virtual particles in the vacuum are not in mass-shell. In local inertial coordinate 
systems $(\bar r,\bar t)$, where gravitational field is absent, the variations of their 
energy $({\cal E}_\circ)$ are described by the Heisenberg uncertainty relationship:
\begin{equation}
\Delta \bar t\Delta {\cal E}_\circ \simeq 1.
\label{h1}
\end{equation}
In arbitrary coordinate systems, the equivalence principle tells us that the Heisenberg 
uncertainty relationship (\ref{h1}) are 
unaffected by the presence of a gravitational field:
\begin{equation} 
\Delta t\Delta {\cal E}\simeq 1.
\label{h2}
\end{equation}
 
These relationships (\ref{h1},\ref{h2}) and the gravitational time dilation between 
local inertial and arbitrary 
coordinate systems $\Delta \bar t = g^{1\over2}(r)\Delta t$ lead us to obtain:
\begin{equation}
\Delta {\cal E}= g^{1\over2}(r)\Delta {\cal E}_\circ,
\hskip0.2cm {\cal E}=g^{1\over2}(r){\cal E}_\circ,
\label{shift}
\end{equation}
where $g^{1\over2}(r)$ (see Eq.(\ref{sg})) is the gravitational red-shift factor.
This indicates that vacuum energy-level ${\cal E}$ and its width 
$\Delta {\cal E}$ of virtual particles are gravitationally red-shifted from corresponding 
vacuum energy-level ${\cal E}_\circ$ 
and its width $\Delta {\cal E}_\circ$ of virtual particles in the absence of
gravitational field. This difference between ${\cal E} $ and ${\cal E}_\circ$ is originated from 
gravitational field interacting with vacuum energy (${\cal E}_\circ$) of virtual particles
in the vacuum. The vacuum gains the energy 
${\cal E}-{\cal E}_\circ <0$ from gravitational field. This result (\ref{shift}) 
implies that the 
vacuum energy ${\cal E}$ varies from one spatial point $(r_1)$ to another $(r_2)$,
\begin{equation}
\delta {\cal E}={\cal E}(r_2)-{\cal E}(r_1)\not=0.
\label{variation}
\end{equation}
This could be in principle examined by measuring the Casimir effect at different altitudes 
above the Earth. 
 
At two different altitudes $r_2$ and $r_1$ above the Earth, Eq.(\ref{shift}) implies 
the Casimir energy $\delta {\cal E}_c$ (\ref{casimir}) should be modified by the gravitation field 
of the Earth in the following way:
\begin{equation}
|\delta {\cal E}_c(r_2)|\!=\!\left({g(r_2)\over g(r_1)}\right)^{1\over2}|\delta{\cal E}_c(r_1)|, 
\hskip0.2cm 
g(r)\!=\!1\!-\!{2M_\oplus\over r},
\label{fermi}
\end{equation}
where $M_\oplus$ is the mass of the Earth. This indicates that $|{\cal E}_c(r_2)|>|{\cal E}_c(r_1)|$ 
for $r_2>r_1$. Given $r_1=r_\oplus$, ${M_\oplus\over r_\oplus}\simeq 7.1\cdot 10^{-10}$ and 
$\Delta r=r_2-r_1=10^6$cm, we obtain
\begin{equation}
|{\cal E}_c(r_2)|\simeq (1+O(10^{-12}))
|{\cal E}_c(r_\oplus)|.
\label{exp}
\end{equation}
Test of this very small energy-gain, modifying the Casimir energy and force, 
seems to be very difficult for current experiments. In the Newtonian regime 
$\min(r_2,r_1)\gg 2M$, 
and $\delta r=r_2-r_1\ll \min(r_2,r_1)$, the variation of the vacuum energy Eq.(\ref{variation}) 
is proportional to $M\delta r/r^2\ll 1$, in practice, too small to be tested.

However, such vacuum energy variations could be enormous, in a gravitational collapse 
approaching to the formation of black hole's horizon. To analyze this, at the 
first we simply model the gravitational collapse as a massive 
star of mass $M$ and radius $R$ undergoing a spherical collapse. 
In the gravitationally collapsing process, the surface of the star moves 
inwards $\delta R>0$ from 
\begin{equation}
\noindent\mbox{the step 1: the radius star} = R+\delta R,
\label{gs1}
\end{equation}
to
\begin{equation}
\noindent\mbox{the step 2: the radius star} = R,
\label{gs2}
\end{equation}
in the time interval $\delta t$. At the step (\ref{gs1}), 
$M$ is the mass distributed 
in the sphere $r<R+\delta R$ of the volume ${4\pi\over 3}(R+\delta R)^3$;
$M'$ is the mass distributed in the sphere $r<R$ of the volume ${4\pi\over 3}R^3$.
$\delta M=M-M'$ is the mass distributed in the spherical shell $R<r<R+\delta R$
of the volume $4\pi R^2\delta R$. 
At the step (\ref{gs2}), the mass $\delta M$ 
falls into the sphere $r\le R$, and total mass $M$ distributed within the sphere $r\le R$. 
Assuming that the matter density is of uniform distribution in this 
gravitational collapse process, we can compute $\delta M$ 
\begin{equation}
\delta M=M(1-{R^3\over (R+\delta R)^3}).
\label{deltam}
\end{equation}
Such a gravitational collapse process $\delta R/\delta t$ can be described by the equation\cite{scoll}
\begin{eqnarray}
\delta t &=&-{2Mh(R)\over g(R)\sqrt{h^2(R)-g(R)}}\delta R,\label{coll}\\
h(R)&=& 1-{2M\over4R}.
\nonumber
\end{eqnarray}

At the second, we study how vacuum energy varies in this gravitational collapse process.
In the absence of gravitational field (inertial frame), we introduce the surface 
vacuum-energy on the surface $r=R$ of the area $4\pi R^2$: 
\begin{equation}
{\cal E}_\circ^s\simeq 4\pi R^2\Lambda^3_p,
\label{s1}
\end{equation}
where $\Lambda_p=\sqrt{\hbar c\over G}=1$ is the Planck scale. In the presence of 
gravitational field, the surface vacuum energy (\ref{s1}) is modified according to 
Eq.(\ref{shift})
\begin{equation}
{\cal E}^s=g^{1\over2}(R){\cal E}_\circ^s.
\label{s2}
\end{equation}
At the step (\ref{gs1}) of the gravitational collapse, 
the surface vacuum-energy on the surface $r=R$ is,
\begin{equation}
{\cal E}'_s=(1-{2M'\over R})^{1\over2}{\cal E}_s^\circ.
\label{deltae1}
\end{equation}
While, at the step (\ref{gs2}) of the gravitational collapse, 
the surface vacuum-energy on the surface $r=R$ is,
\begin{equation}
{\cal E}_s=(1-{2M\over R})^{1\over2}{\cal E}_s^\circ. 
\label{deltae2}
\end{equation}
We find that $M'$ in Eq.(\ref{deltae1}) is altered to $M$ in Eq.(\ref{deltae2}), since the
mass $\delta M$ in the spherical shell $R<r<R+\delta R$ falls into the sphere $r<R$.
The vacuum-energy variation in this gravitational collapse process from the step 
(\ref{gs1}) to the step (\ref{gs2}) is 
\begin{equation}
\delta {\cal E}={\cal E}_s - {\cal E}'_s<0,
\label{deltae3}
\end{equation}
which indicates the vacuum gains energy from gravitational field. 

\begin{figure}
\includegraphics[width=\hsize]{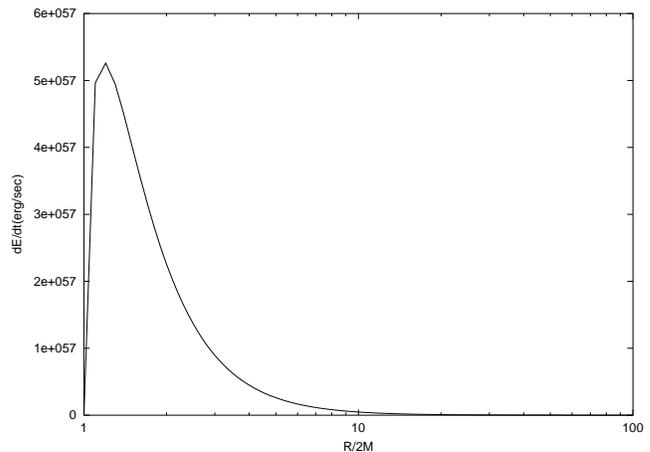}
\caption{The rate of energy release $\delta {\cal E}/\delta t$ (erg/sec) as a function of the radius $R$
in unit of $2M$.} 
\label{rate}
\end{figure}

By using Eqs.(\ref{coll}) and (\ref{deltae3}), we compute the rate of energy gain 
$\delta {\cal E}/\delta t$ in the spherical shell $4\pi R^2dR$ that the surface of the collapsing 
star sweeps in the time interval $\delta t$.
Given the initial condition that at the moment 
$t_\circ=0$ of starting the collapsing process, the radial size of the collapsing star 
$R_\circ=100(2M)$ and star's mass $M=10M_\odot$, we compute the rate of vacuum-energy 
variation (gain) $\delta {\cal E}/\delta t$, plotted in Fig.(\ref{rate}) as a function of $R$ 
in the unit of $2M$. 
The result shows that the rate $\delta {\cal E}/\delta t$ rapidly increases to 
$10^{57}$erg/sec, 
as the surface $R(t)$ of the collapsing star 
moves, almost in the speed of light, inwards to the horizon. Whereas, in the vicinity of the 
horizon, the collapsing process becomes slow and the rate decreases.

Due to this vacuum-energy gain $\delta {\cal E}$ (\ref{deltae3}), 
vacuum states become energetically unstable, have to spontaneously undergo a quantum
transition to lower energy states via quantum-field fluctuations. 
This is exactly analogous to the phenomenon of the Casimir effect. As a consequence, the 
vacuum-energy $\delta {\cal E}$ (\ref{deltae3}) 
gained from gravitational field must be released and deposited in the region 
from $r=2M$ extending to $r=R_\circ$.

Which process of 
quantum transition releases this vacuum-energy $\delta {\cal E}$ (\ref{deltae3})? 
One of possibilities is spontaneous photon emission, analogous to the
spontaneous photon emission taking place in the atomic physics. Such a spontaneous 
photon emission can be induced by the four-photon interacting vertex in the theory of 
Quantum Electromagnetic Dynamics (QED). 
The rate of the quantum transition must be proportional to $\sim\alpha^4\omega$, 
where $\omega$ is the characteristic energy of the process in the time interval 
$\delta t$. For high-energy $\omega\sim m_e$, where $m_e$ is the electron mass,
the rate is very fast\cite{add}. As shown in Fig.(\ref{rate}), 
the spatial density of vacuum energy release can be very large, as the collapsing process 
approaching to the formation of black hole's horizon $R=2M$. The energy of photons 
spontaneously emitted can be larger than the energy threshold $2m_e$, so that electron and 
positron pairs are produced. These pairs, on the other hand, annihilate into
two photons. As a consequence, a dense and energetic plasma of photons, electron and 
position pairs, called ``{\it dyadosphere}'' \cite{prx} or ``{\it fireball}'' in 
literatures\cite{piran}, could be formed. 
Using the rate of collapsing $\delta R/\delta t$ (\ref{coll}) and the 
rate $\delta {\cal E}/\delta t$ of vacuum-energy release process (Fig.(\ref{rate})), 
we can obtain that total amount energy:
\begin{equation}
E_{\rm total}=\int_{2M}^{R_\circ}d {\cal E} \lesssim {3\over 2\pi} M\simeq 8.6\cdot 
10^{54}{\rm erg},
\label{rdem}
\end{equation}
is released in a very short time, about a second for the collapsing process 
from $R=R_\circ$ to $R=2M$. These qualitatively agree to the characteristic of 
energetic sources for gamma ray bursts.

The total energy release Eq.(\ref{rdem}) is less than the maximum variation 
$\Delta E_g={1\over2}M$ of gravitation energy(potential $-{M\over r}$) in the collapse 
prosess from $r\sim\infty$ to $r=2M$. The model and approach that we adopt only 
to illustrate the idea and scenario, is over simplified for quantitative results and 
will be elaborated in future work. Note added: after finishing this paper, the author was
interested in reading the paper by I.Yu Sokolov (Phys. Lett.A, v.223, p.163, 1996), where 
conducting electron gas is used as boundary conditions for computing the vacuum energy to 
discuss possible huge output of cosmic energy accounting for Quasars.

\newpage

\end{document}